\documentclass[12pt]{article}
\usepackage{graphicx}
\usepackage{cite}
\usepackage{color}
\usepackage{sidecap}
\newcommand{\bm}[1]{\mbox{\boldmath $#1$}}
\newcommand{\abs}[1]{\left| #1\right|}
\newcommand{\fnd}[2]{\frac{\textstyle #1}{\textstyle #2}}
\newcommand{\fndrs}[4]{\fnd{\raisebox{#1}{$#2$}}{\raisebox{#3}{$#4$}}}
\newcommand{\dissum}[2]{\displaystyle \sum_{#1}^{#2}}
\newcommand{\figline}[1]{(\begin{picture}(12,0)(0,0)%
\end{picture})}
\long\def\symbolfootnote[#1]#2{\begingroup
\def\thefootnote{\fnsymbol{footnote}}
\footnote[#1]{#2}\endgroup}

\begin{document}
\noindent
{\Large\bf The quark-antiquark spectrum from upside down$^{\star}$}
\symbolfootnote[0]{\footnotesize \hspace{-10pt}$^{\star}$Talk
given by Eef van Beveren}\\ [20pt]
Eef van Beveren$^{\it a}$ and George Rupp$^{\it b}$\\ [10pt]
{\small $^{\it a}$Centro de F\'{\i}sica Computacional,
Departamento de F\'{\i}sica,
Universidade de Coimbra,\\ P-3004-516 Coimbra, Portugal}\\
{\small $^{\it b}$Centro de F\'{\i}sica das Interac\c{c}\~{o}es Fundamentais,
Instituto Superior T\'{e}cnico,\\
Universidade T\'{e}cnica de Lisboa, Edif\'{\i}cio Ci\^{e}ncia,
P-1049-001 Lisboa, Portugal}\\ [30pt]
{\small {\bf Abstract.} We argue that the spectra of
quark-antiquark systems should better be studied from higher
radial excitations and, in particular, from configurations
with well-defined quantum numbers, rather than from ground states
and lower radial excitations,
the most suitable system being charmonium.}
\vspace{20pt}

In the Resonance-Spectrum Expansion (RSE) \cite{IJTPGTNO11p179},
which is based on the model of Ref.~\cite{PRD21p772},
the meson-meson scattering amplitude is given by an expression
of the form
(here restricted to the one-channel and one-delta-shell case)
\begin{equation}
\bm{T}(E)=\left\{ -2\lambda^{2}\mu p
j_{\ell}^{2}\left( pr_{0}\right)
\dissum{n=0}{\infty}
\fndrs{5pt}{\abs{g_{nL(\ell )}}^{2}}{-2pt}{E-E_{nL(\ell )}}
\right\}
\bm{\Pi}(E)
\;\;\; ,
\label{Tamplitude}
\end{equation}
where $p$ is the center-of-mass (CM) linear momentum,
$E=E(p)$ is the total invariant two-meson mass,
$j_{\ell}$ and $h^{(1)}_{\ell}$ are
the spherical Bessel function and Hankel function of the first kind,
respectively, $\mu$ is the reduced two-meson mass,
and $r_{0}$ is a parameter with dimension mass$^{-1}$,
which can be interpreted as the average string-breaking distance.
The coupling constants $g_{NL}$,
as well as the relation between $\ell$ and $L=L(\ell)$,
were determined in Ref.~\cite{ZPC21p291}.
The overall coupling constant $\lambda$,
which can be formulated in a flavor-independent manner,
represents the probability of quark-pair creation.
The dressed partial-wave RSE propagator
for strong interactions is given by
\begin{equation}
\bm{\Pi}_{\ell}(E)=\left\{
1-2i\lambda^{2}\mu pj_{\ell}\left( pr_{0}\right)
h^{(1)}_{\ell}\left( pr_{0}\right)
\dissum{n=0}{\infty}
\fndrs{5pt}{\abs{g_{NL}}^{2}}{-2pt}{E-E_{NL}}
\right\}^{-1}
\;\;\; .
\label{propagator}
\end{equation}

In Ref.~\cite{ARXIV08111755} we have studied an intriguing property
of the propagator (\ref{propagator}),
namely that it vanishes for $E\to E_{NL}$.
Here, we will concentrate on the fact that
the scattering amplitude (\ref{Tamplitude}) is independent
of the way quark confinement is introduced,
as only the confinement spectrum $E_{NL}$ appears
in expressions (\ref{Tamplitude}) and (\ref{propagator}).
Hence, whatever one's preferred mechanism for confinement,
the scattering amplitude only depends on the resulting spectrum.
As a consequence, one merely has to deduce from experiment
a suitable set of values of $E_{NL}$
and then try to guess the corresponding dynamics for confinement.

However, in the recent past we have found that
analysing experimental data is far from trivial.
For instance, the expressions (\ref{Tamplitude}) and (\ref{propagator})
may also lead to dynamically generated resonances,
like the light scalar-meson nonet \cite{ZPC30p615},
or the $D_{s0}(2317)$ \cite{PRL91p012003},
which do not stem directly from the confinement spectrum.
Furthermore, the mountain-shaped threshold enhancements
in plots of events versus invariant mass for particle production
can easily be mistaken for resonances \cite{PRD80p074001}.
Also, to properly analyse certain hadronic decay modes,
one has to turn the resulting data upside down,
so as to find the true quarkonium resonances and threshold enhancements
\cite{PRL105p102001}, instead of erroneously classifying
the leftovers as unexpected new resonances \cite{PRL95p142001}.
Moreover, in studying resonances from production processes,
one has no control over their quantum numbers \cite{PRD80p092003}.

Our simple formulas (\ref{Tamplitude}) and (\ref{propagator})
for meson-meson scattering are certainly not good enough
for a detailed description of production processes,
but must be adapted in order to account
for, at least, the threshold enhancements \cite{AP323p1215}.
However, the precise dynamics of production processes is still far from
being fully understood.
Nevertheless, for the low-lying part of the spectra
we may deduce some properties without too much dependence
on a specific confinement spectrum.
We found that meson loops, which are properly accounted for
in expressions (\ref{Tamplitude}) and (\ref{propagator}),
have most influence on the mass shifts of the ground states.
Consequently, upon deducing a confinement spectrum
from the lowest-lying states,
one is urged to seriously consider the meson loops \cite{PRD21p772}.

Threshold enhancements are more conspicuous for
sharp thresholds, i.e., when the involved particles
have small widths, rather than for diffuse thresholds,
concerning decay products that have considerable
widths themselves \cite{ARXIV10085100}.
The latter phenomenon tends to happen higher up in the spectrum.
There, we may expect a smoothened-out pattern of overlapping
broad threshold enhancements.
Therefore, higher radial excitations of quarkonium resonances
can more easily be disentangled from other enhancements.
The disadvantage is that any confinement mechanism predicts,
for higher excitations,
abundantly many states of the $q\bar{q}$ propagator,
with a variety of different quantum numbers.

Now, in order to avoid a large number of partly overlapping
resonances, one best studies resonances obtained
in electron-positron annihilation, which process is dominated
by vector quarkonia.
But this is not the full solution for cleaning up the data,
since in the light-quark sector one has
nonstrange and strange $q\bar{q}$ combinations with comparable spectra,
which will come out on top of each other,
besides possibly significant mixing
of isoscalar $n\bar{n}$ ($n=u,d$) and $s\bar{s}$ states.
Moreover, decay channels involving kaons are common to
both $n\bar{n}$ and $s\bar{s}$ resonances.
Actually, the only system with a sufficient number of
established states to find evidence
(see Table 3 of Ref.~\cite{EPJA31p468})
for a regular level splitting of about 380 MeV
is given by the radially excited $f_{2}$ mesons.
A way out is to study a well-isolated system, with just one
set of quantum numbers, like vector $c\bar{c}$ states,
which can be produced in $e^{+}e^{-}$ annihilation.
Once the spectrum of vector $c\bar{c}$ is well established,
one can with some confidence apply its properties to other spectra.
Unfortunately, the well-established $J^{PC}=1^{--}$ $c\bar{c}$ spectrum
anno 2010 still consists of
$J/\psi$, $\psi (2S,3S,4S)$, and $\psi (1D,2D)$ only.

\begin{SCfigure}[50][htbp]
\centering
\begin{tabular}{|c|}
\hline\\ [-11pt]
\includegraphics[height=220pt]{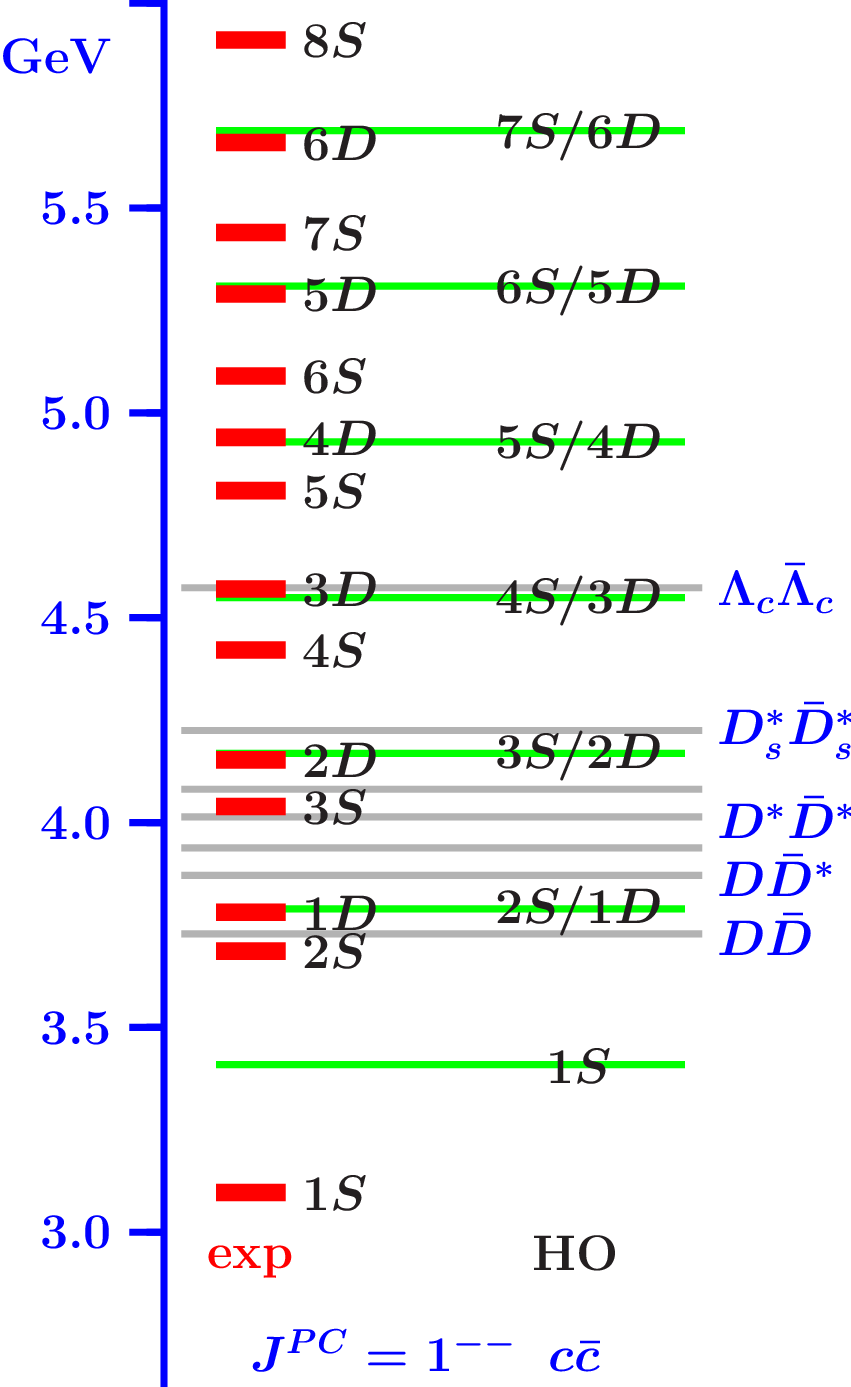}\\
\hline
\end{tabular}
\caption[]{\small
The higher charmonium vector states
({\color{red}exp}) as extracted by us from data:
(i) the $\psi (3D)$ \cite{PRL105p102001},
in BABAR data \cite{PRL95p142001}
on $e^{+}e^{-}\to J/\psi\pi^{+}\pi^{-}$;
(ii) the $\psi (5S)$ and $\psi (4D)$ \cite{ARXIV10053490},
in data obtained by the Belle Collaboration
on $e^{+}e^{-}\to\Lambda_{c}^{+}\Lambda_{c}^{-}$ \cite{PRL101p172001},
$D^{0}D^{\ast -}\pi^{+}$ \cite{PRD80p091101},
and $D^{0}D^{\ast -}\pi^{+}$ \cite{PRL100p062001},
as well as in the missing signal of Ref.~\cite{PRL95p142001},
and in further BABAR data on $D^{\ast}\bar{D}^{\ast}$
\cite{PRD79p092001};
(iii) the $\psi(5S)$, $\psi(4D)$, $\psi(6S)$, and $\psi(5D)$
\cite{EPL85p61002}, in the data of Ref.~\cite{PRL101p172001};
(iv) the $\psi(3D)$, $\psi(5S)$, $\psi(4D)$, $\psi(6S)$, and $\psi(5D)$
\cite{ARXIV09044351},
in new, preliminary BABAR data \cite{ARXIV08081543}
on $e^{+}e^{-}\to J/\psi\pi^{+}\pi^{-}$;
(v) the $\psi(7S)$, $\psi(6D)$, and $\psi(8S)$ \cite{ARXIV10044368},
in data from BABAR on $D^{\ast}\bar{D}^{\ast}$ \cite{PRD79p092001}.
We also indicate the level scheme
as predicted by pure HO confinement (HO) \protect\figline{0 1 0}.
Meson and baryon loops shift the $D$ states a few MeV down/up,
whereas the $S$ states shift 100--200 MeV downwards.
For completeness, we also indicate the levels
of the sharp, low-lying meson-meson and baryon-baryon thresholds
\protect\figline{0.7 0.7 0.7} of the channels
$D\bar{D}$, $D\bar{D}^{\ast}$, $D_{s}\bar{D}_{s}$,
$D^{\ast}\bar{D}^{\ast}$, $D_{s}\bar{D}_{s}^{\ast}$,
$D_{s}^{\ast}\bar{D}_{s}^{\ast}$, and $\Lambda_{c}^{+}\Lambda_{c}^{-}$.
}
\label{HigherCharmonium}
\end{SCfigure}

In the following, we will concentrate on a specific choice
for confinement, namely the harmonic oscillator (HO), though not
so much the corresponding potential or geometry (anti-De-Sitter
\cite{NCA80p401}), but just the HO spectrum that follows from these
approaches.  For vector $c\bar{c}$ systems, one has a single $^{3\!}S_1$
ground state, and radial excitations, which can be either $^{3\!}S_1$
or $^{3\!}D_1$. In the HO spectrum, $^{3\!}S_1$ states with radial quantum
number $n$ and $^{3\!}D_1$ states with $n\!-\!1$ are degenerate.
However, due to the interaction generated by the meson loops,
the poles associated with the resonances repel each other in such a way
that one is subject to a small mass shift,
whereas the other shifts considerably more and downwards.
Higher up in the $c\bar{c}$ spectrum,
the mass shift of the lower pole,
which is dominantly $^{3\!}S_1$, becomes of the order of 150--200 MeV,
whereas the higher pole, mostly $^{3\!}D_1$,
acquires a central resonance position that is only
a few to at most about 50 MeV away from the HO spectrum \cite{PRD21p772}.
In Ref.~\cite{ARXIV10044368} we found evidence, in data
obtained by the BABAR Collaboration \cite{PRD79p092001}, for further
charmonium states, viz.\ $\psi(5S,6S,7S,8S)$ and $\psi(3D,4D,5D,6D)$,
which confirm the above observation.
In Fig.~\ref{HigherCharmonium} we display the resulting
spectrum for vector charmonium.

We can conclude from Fig.~\ref{HigherCharmonium}
that our guess of an HO spectrum
for vector $c\bar{c}$ states, with a radial level spacing of 380 MeV
\cite{PRD27p1527}, seems to work well in view of the data.
Furthermore, we may observe the advantage of studying
the $c\bar{c}$ vector spectrum from above, where the pattern
of dominantly $S$ and dominantly $D$ states
becomes rather regular.
\vspace{10pt}

{\bf Summarizing},
we have shown that the $c\bar{c}$ confinement spectrum,
which underlies scattering and production of multi-meson systems
containing open-charm pairs,
can best be observed by starting from
higher radial excitations of vector charmonium
in electron-positron annihilation.
Moreover, we have shown that a constant radial level splitting
of about 380 MeV is consistent with light and heavy meson spectra.
\vspace{10pt}

We wish to thank Jorge Segovia for very fruitful discussions.
One of us (EvB) thanks the organizers
of the Mini-Workshop "Bled 2010", for arranging a very pleasant
setting for discussion and exchange of ideas.
This work was supported in part by the {\it Funda\c{c}\~{a}o para a
Ci\^{e}ncia e a Tecnologia} \/of the {\it Minist\'{e}rio da Ci\^{e}ncia,
Tecnologia e Ensino Superior} \/of Portugal, under contract
CERN/\-FP/\-109307/\-2009.

\newcommand{\pubprt}[4]{#1 {\bf #2}, #3 (#4)}
\newcommand{\ertbid}[4]{[Erratum-ibid.~#1 {\bf #2}, #3 (#4)]}
\def\AP{Ann.\ Phys.}
\def\EPJA{Eur.\ Phys.\ J.\ A}
\def\EPL{Europhys.\ Lett.}
\def\IJTPGTNO{Int.\ J.\ Theor.\ Phys.\ Group Theor.\ Nonlin.\ Opt.}
\def\NCA{Nuovo Cim.\ A}
\def\PRD{Phys.\ Rev.\ D}
\def\PRL{Phys.\ Rev.\ Lett.}
\def\ZPC{Z.\ Phys.\ C}

\end{document}